# AN IMPROVED ACCELERATED FRAME SLOTTED ALOHA (AFSA) ALGORITHM FOR TAG COLLISION IN RFID

Tanveer Ahmad[1] and Yan tian feng[2]

[1]School of Electronic and Information Engineering, Lanzhou Jiaotong University
Lanzhou, China
Email: - hons_graduate@yahoo.com
[2]School of Electronic and Information Engineering, Lanzhou Jiaotong University
Lanzhou, China
Email: - yantianfeng@163.com

## ABSTRACT

*The efficiency of tag identification in an RFID (radio identification system) can be low down due to the tag collision problems. The tag collision problem occurs when a reader try to read multiple tags in an interrogation zone. As a result, the reader doesn't identify the tag correctly. That causes a loss of information or data interference. To solve such kind of issues a series of ALOHA based algorithm and Binary search algorithm have been proposed. The most simple, popular and good giving performance algorithms are ALOHA based anti-collision algorithms. In this paper we present a new variation in Accelerated slotted Aloha (AFSA). Our proposed Algorithm by using the bitmaps and avoids wastages in bit times due to idleness and collided slots, reduce the tag reading time. The simulation result shows that AFSA can significantly reduce the average tag reading time with respect to the base protocols and achieve high tag reading rates under both static and mobile settings.*

## 1. INTRODUCTION

Traditional bar-coding technology is an economical solution for AIDC (Automatic Identification data Collection). By rapid usage we find some limitations on it. We need to scan each bar code item individually, that limiting the scanning speed and increase the cost of process by the use of manual labour. The manual process of scanning also increases the probability of human error. As a result engineers face a challenge to develop a new technology that ensures the efficiency and reliability of identification process. For the reason RFID technology has been making inroads in AIDC applications.

By the use of RF signal RFID technology make it possible to identify and collect the object data quickly and efficiently. The analysis process shows that RFID is much similar to bar-coding technology, but the biggest advantage of RFID is that it does not rely on the line of sight [1]. RFID is an emerging technology making ubiquitous identification possible. High speed moving objects and recognition of multiple tags simultaneously can be identified by RFID. Combined with the internet, telecommunication and other technologies, RFID technology can achieve track and share object information world-widely.

The main advantages of RFID over barcodes are [3] [4]

- Don't require direct line of sight.
- Rugged then bar codes.
- Fastest reading speed
- Read/ write capacity as design allow.
- No need to attach outside the products.
- Greater flexibility.
- Higher data storage
- Increased data collection through put.
- Greater accuracy of data.
- Durability. (RFID tags still working if it become dusty)

Object identification dilemma requires the simultaneous recognition of multiple objects/ tags reliably and minimal user interaction. The key problem of RFID is its read rate and recognition speed. In RFID system there are two kinds of interference: 1. Tag interference and 2. Reader interference. At this stage our discussion is limited to tag interference

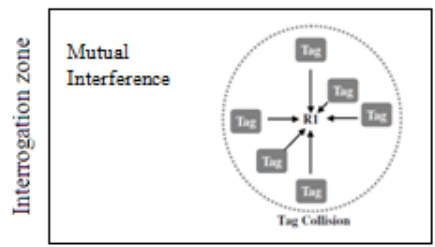

Figure. 1 Problems in transmission of data from
many individual tags to reader.

In a dense RFID environment when multiple tags try to communicate with a reader simultaneously in its interrogation zone, the tags will collided and no more communication is possible. This is called tag collision or tag-to-tag interference. A reader cannot identify multiple tags because of signal collision. The tag in this state is in dead zone called as tag jamming state. Its results in surplus of energy, bandwidth as well as increase identification delay [2]. Figure 1 is a pictorial diagram of tag interference.

The rest part of this paper is prepared as follows: Section II describes the components and review of RFID technologies and their short comparison based on their operating principles. In Section III we present a short description of current Anti-collision protocols. Section IV describe our projected algorithm based on Accelerated frame slotted Aloha (AFSA). Section V concludes the paper.

## 2. BACKGROUND

Before moving to Anti-collision protocols discussion this section shows the components and the working/ operation of RFID technology.

## 2.1. Components of RFID System

A typical RFID system consist of three components as illustrated in figure 2 [3][4].

- One or more RFID tags, called as transponder, that affix to the object to be recognized. Tags may be active or passive. Active tags have its own battery source but passive tag don't require power source. It can be powered by the reader. Table 1 shows the common difference between passive and active tags.

- A reader or transceiver is a power full device with built in memory. RFID reader has two interfaces. The first one is RF interface that communicate with tags in their interrogation zone and the second one is a communication interface for interacting with server

- Finally an Application/ data processing sub-system, which may be an application/ data base that depend upon the application.

Table 1.  Difference between Active and Passive Tags.

| Factors | Tag Types | |
| --- | --- | --- |
| | **Active** | **Passive** |
| **Power** | Internal | External |
| **Life** | Low | High |
| **Transmitter** | Yes | No |
| **Cost** | Expensive | Cheap |
| **Read Range** | (60-300) feet | Up to 30 feet |
| **Data Storage** | High | Less |
| **Size** | Slightly Bulky (Due to battery) | Small/ light weight |
| **Maganetic Field Strength** | Low | High |

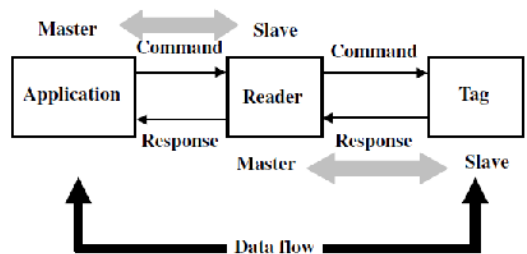

Figure. 2 Master-Slave principle between application, reader, and tag.

## 2.2. Communication Principal

Magnetic or electromagnetic coupling is used for RFID communication. The difference between these two system lies in their operating field i.e. near or far field [2][5][6][7]. Table II presents a comparison between magnetic and electromagnetic coupling system.

TABLE II.    MAGNETIC AND ELECTROMAGNETIC COUPLING

| Magnetic coupled system | Electromagnetic coupled system |
|---|---|
| Operate in LF or HF band | Operate in UHF and microwave band |
| Passively Operate | Active |
| Transformer type coupling | Backscatter Coupling |
| Using Amplitude modulation | Using RF power transmission |
| Low Range | High Range |

## 3. EXISTING ANTI-COLLISION PROTOCOLS

There are several methods to handle the multi-tag interference. The most common algorithms to reduce collision are Probabilistic and deterministic. The deterministic schemes are ALOHA based anti collision protocols while deterministic refer to binary tree search. ALOHA based algorithms are very popular in industry. It schedules the response of tags at random time that leads to low probability of collision. As for this paper our discussion is not towards Anti-collision protocol discussion. So after a brief description we move towards our proposed algorithm. Figure    3 shows the Taxonomy of existing anti-collision protocols.

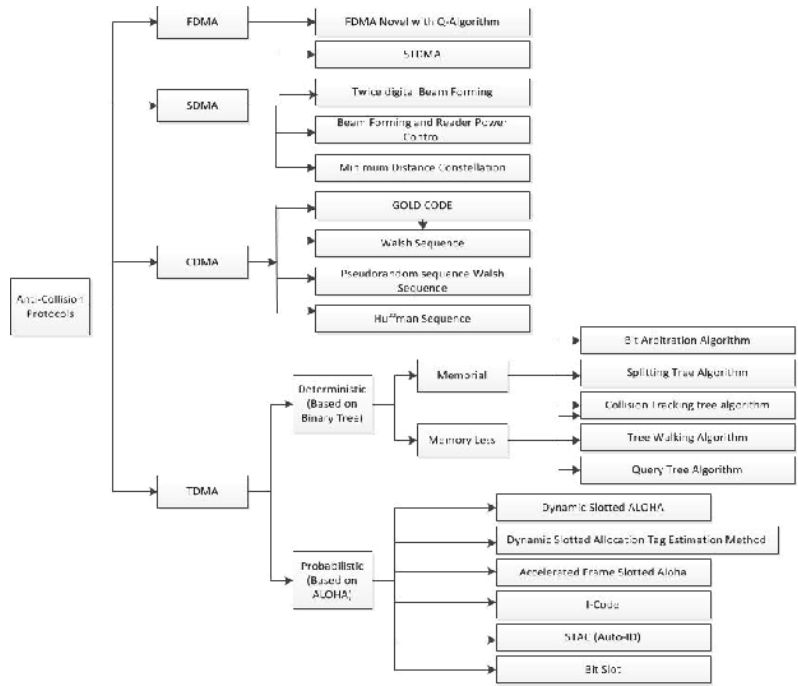

Figure. 3 Taxonomy of Anti-collision Methods in RFID system

## 4. DESCRIPTION OF THE PROPOSED ALGORITHM

We suppose the collision limit 900 MHz UHF RFID system. One tag transmits 64-bit identifier (EPC) and 16-bit CRC with 4μs time duration. A reader sends a query to tag by '0', '1', and Null. Reader time duration is 12μs [8] and it uses 900MHz continuous waves (CW) to power the tag. The proposed algorithm is a variation in AFSA, which has been implemented in MATLAB for getting simulation result. The operation will accomplish in five different phases.

1. In advertisement phase reader broadcast frame size (N) to all tags. Tag generates a random number, which allows the tag to participate in the current round if it gets remainder zero. In case of participation tag generate another random number from 0 to N to get the slot.

2. In Reservation phase n bit sequences are sending by tag in their chosen slot. For n there is $2^n$ promising n bit sequence. A tag arbitrarily picks one of these $2^n$ sequences to transmit these sequences to reader by chosen slot.

3. In third phase of reservation summary a tag send summary of reservation to tag. Suppose N=4 the bitmap 1001 shows the tag chosen slot 1 & 4 as successfully reserved. In this case reader did not hear at slot 2 & 3 because of (i) collision I these slots or (ii) idle slots.

4. The next phase is data transmission phase in which the tag who find his slot successfully send the identifier and other associated data to reader.

5. Finally reader sends the acknowledgement to tag in the shape of bit. '1' represents successful transmission and vice versa. Figure 4.

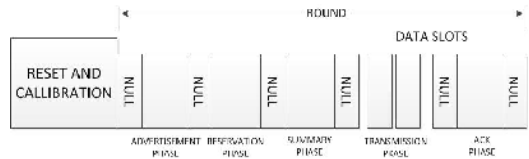

Figure. 4 Rounded Phases of AFSA

For the tag estimation we choose the approaches presented in [8] because of its simplicity. Let K tags are being estimated by the reader after completion of a round. As stated above each tag randomly transfers n-bit sequence in the selected slot. Let T be the total time of round $T_{ad}$, $T_R$, $T_{su}$, $T_D$, $T_{ACK}$ Table III [8]. Then we can write T as.

$$T= T_{ad} + T_R + T_{su} + T_D + T_{ACK} \text{ ----------------} \qquad (1)$$

Thus the duration of reservation phase is $12.5^*Nn\mu s$, while $T_D$ is $S^*320\mu s$. S indicates the no of successful slots reservation i.e.

$$S=E[R] + E[UC] \text{--------------------} \qquad (2)$$

Where E[R] is average no of truly reserved slots, and E[UC] is the average no of collision slots. Since the tag choose one slot from N, so the values of average no of reserved slots can be written as [8].

$$E[R] = K (1-1/N)^{K-1} \text{ -----------------} \qquad (3)$$

If multiple tags choose one slot and all tags choose the identical bit sequence, then the calculation of average no of slot is

$$E[UC] = N \sum_{i=2}^{K-E[R]} (Probability \ that \ i \ tags \ in \ the \ same \ slot) \frac{1}{2^{n-1}}$$

Let E[U] signify the total no of slots that more then on tag choose.

E [U] = N-E [I]-E[R] --------------- (5)

Where E [I] is total no of idle slots.

E [I] = N (1-1/N) $^K$ ----------------- (6)

TABLE III.    MAGNETIC AND ELECTROMAGNETIC COUPLING

| Symbol | Description |
|--------|-------------|
| $T_{ad}$ | Time of advertisement Phase |
| $T_R$ | Time of Reservation Phase |
| $T_{su}$ | Time of Reservation summary Phase |
| $T_D$ | Time of Data transmission Phase |
| $T_{ACK}$ | Time of Acknowledge Phase |

If $P_j$ the probability that two tags picks the identical n bits sequence in $T_R$. then from equation 4

$$E [UC] \quad E [U]. P_j$$

As $P_j = 1/2^n$     by combining all equation T becomes.

$$T= T_{ad}+12.5Nn+T_{su}+320(E[R]+E[U].P_j)+12.5(E[R]+E [U]. P_j)$$

By minimizing with esteem to n the optimal value of n is

$$n* = 3.32 \log 10 (19.13E[U] /N)$$

In optimal n* calculation some of the collision go undetected during the reservation phase. Figure 5 shows the simulation result that getting after 25 experiments, is a plot of undetected collision for each tag with different n values. The proposed result show that if we have a large no of n bit sequences the undetected collision goes down.

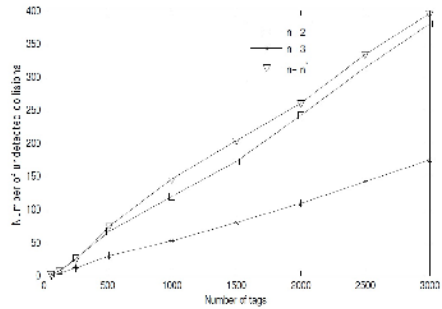

Figure. 5 Number of Undetected Collision vs. N

Figure 6 shows the simulation result of average tag identification time in m sec. by experiment when we use n=2 we achieve better tag reading time as compare to n values for 3 and 4 respectively. The optimal values of n* and n=2 is same in average reading time. The reason is that n* computed values always circulate around 2.

The comparison of AFSA with EDFSA and ASAP shows that AFSA tag identification time is less. Figure 7 gives the average for tag recognition period. The simulation shows that the average tag reading time for EDFSA is 1 ms where as ASAP average identification time is 059ms. Our proposed AFSA have steady performance with an average identification time of 0.49 ms.

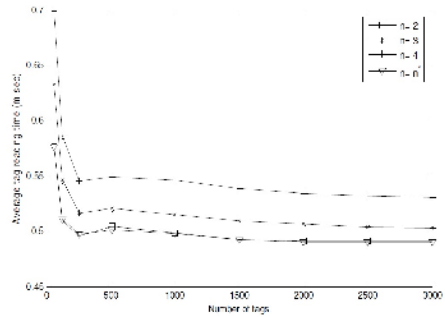

Figure. 6 Average tag identification times vs. N

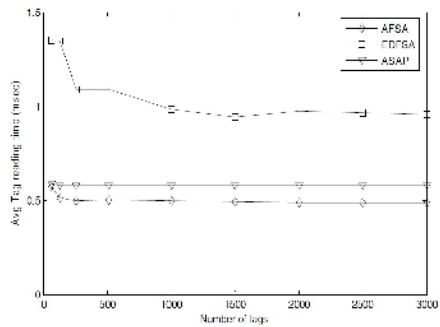

Figure. 7 optimal value n* (AFSA) comparison with EDFSA and ASAP

## 6. Conclusion

The main issue in RFID system is a tag collision. There may be more then one tag in reader working range. When reader queried tags simultaneously it will create a tag collision paradigm. That causes the loss of information in a sophisticated identification system. This paper presents the AFSA protocol for resolving this interference. In this proposed algorithm we try to reduce the tag reading time with the passive use of bitmaps and avoid the surplus in bit selection time. Also the simulation result shows that the proposed algorithm has high efficiency and lower processing time.

## Authors

**Tanveer Ahmad** received his Bachelor degree from COMSATS Institute of Information technology, Pakistan, in 2006. He is a graduate student in Communication and Information System, Lanzhou Jiaotong University, GANSU, Lanzhou, China from 2011. His current interest includes Wireless communication, Mobile network communication and RFID.

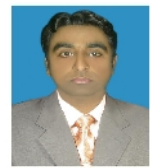

**Yantian feng** is a professor of Lanzhou Jiaotong University, GANSU, Lanzhou, China. His current interest includes Software defined Radio and Radio spectrum monitoring.

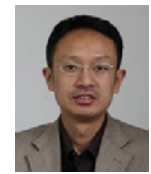